\documentclass[11pt]{article}
\usepackage{graphicx}
\usepackage{hyperref}
\pdfoutput=1
\begin{document}
\title{Magnetocapillary Swimmers}
\author{Maxime Hubert, Galien Grosjean, Yves-Eric Corbisier, \\Geoffroy Lumay, Floriane Weyer, Noriko Obara, \\ Nicolas Vandewalle \\
\\\vspace{6pt} GRASP, Physics Dept., Sart Tilman   \\ University of Li\`{e}ge, B4000 Liège, Belgium}
\maketitle
\begin{abstract}
We present an experiment where three mesoscopic soft ferromagnetic beads are placed onto a liquid surface and submitted to the influence of magnetic fields.  A vertical magnetic field creates a repulsion which counterbalances the capillary attraction. We show that the competition with a second, oscillating field, deforms the structure in a non reciprocal way. As a consequence, the structure is able to swim. This experiment is fully described in a fluid dynamics video attached to this submission.
\end{abstract}
\section{Introduction}
When particles are placed on some liquid-air interface, the deformations of this surface around the particles imply either attractive or repulsive capillary interactions. Self-assembly due to capillary forces is an elegant method for generating 2D mesoscale structures \cite{Whitesides}. Our experiments consider soft ferromagnetic beads placed on a water-air interface. In this configuration, the beads agglomerate due to capillary attraction. This behavior is called the \emph{Cheerios effect} \cite{Vella}. The presence of a vertical magnetic field creates a magnetic dipole into the beads which therefore experience magnetic repulsions \cite{Vandewalle}. A second horizontal oscillating magnetic field induces periodic oscillations of the beads. Thus, the structure created by the beads deforms \cite{Lumay}. \\

\section{Experimental Setup}
The experimental setup is the following. Soft ferromagnetic beads (AISI 52100 alloy) of diameter D=397$\mu$m and D=500$\mu$m are dropped on a water-air interface in a petri dish. This system is placed between two horizontal coils in the Helmholtz configuration which create the vertical magnetic field. The field amplitude lies between 15G and 55G. Indeed, below 15G, the structure collapses. Above this limit, the interdistance between the beads is directly related to the field magnitude \cite{Vandewalle}. Typical interdistances range between one and four bead diameters. A second set of Helmholtz coils, placed perpendicularly to the first set of coils, creates the oscillating field. The applied frequency varies from 1Hz to 8Hz and the amplitude lies between 5G and 40G. \\

\section{Results}

\begin{figure}[!]
\centering
\includegraphics[scale=0.35]{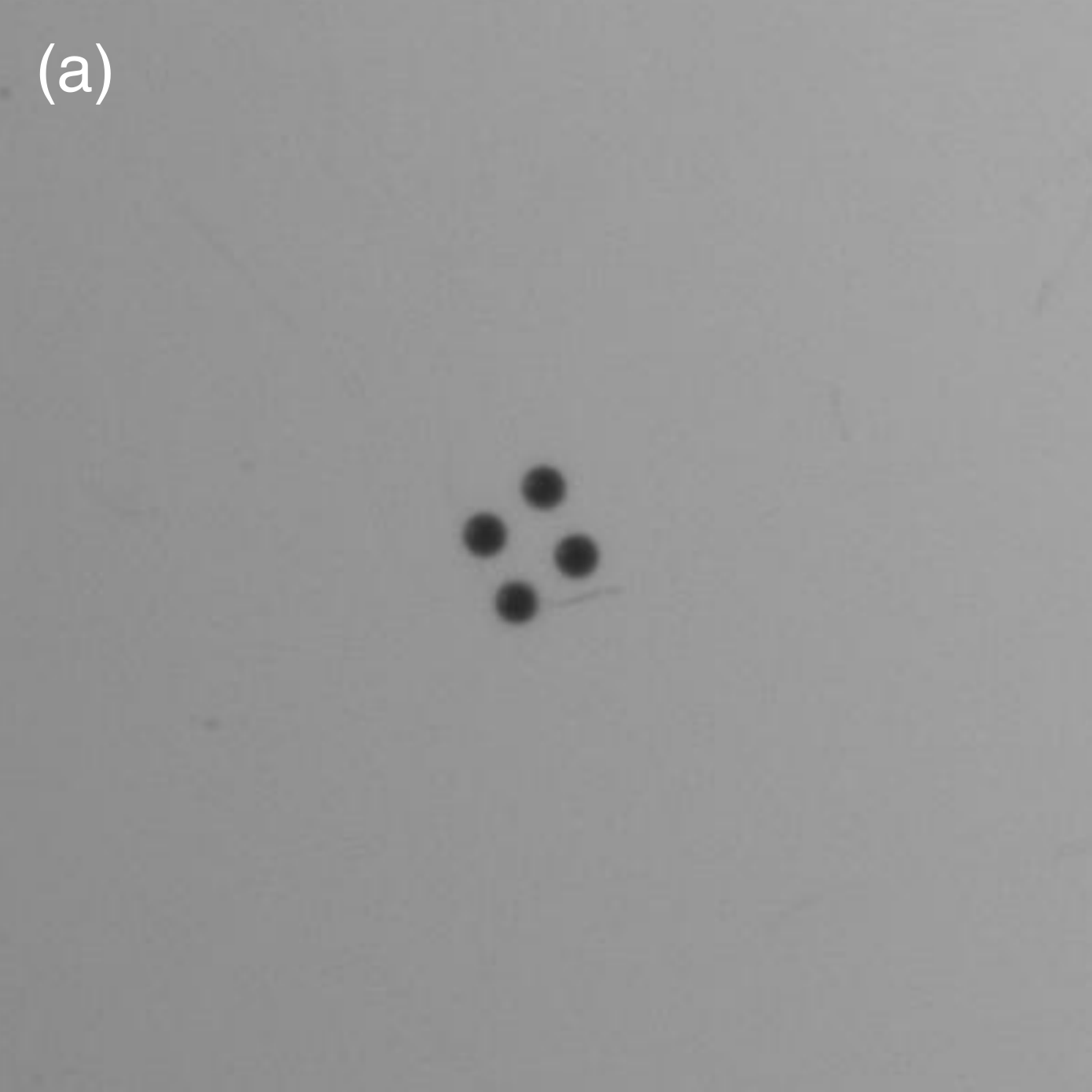} 
\includegraphics[scale=0.35]{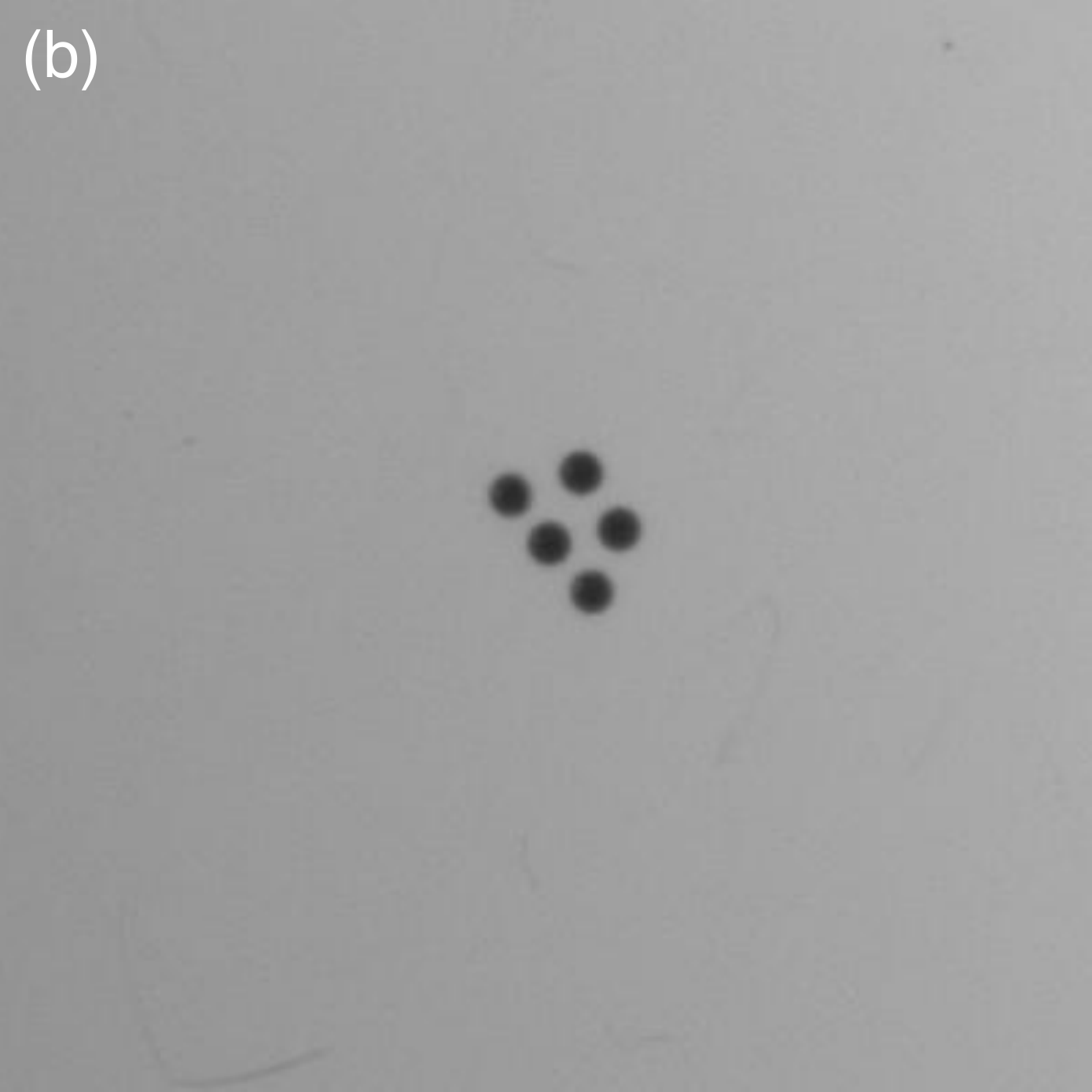} 
\includegraphics[scale=0.35]{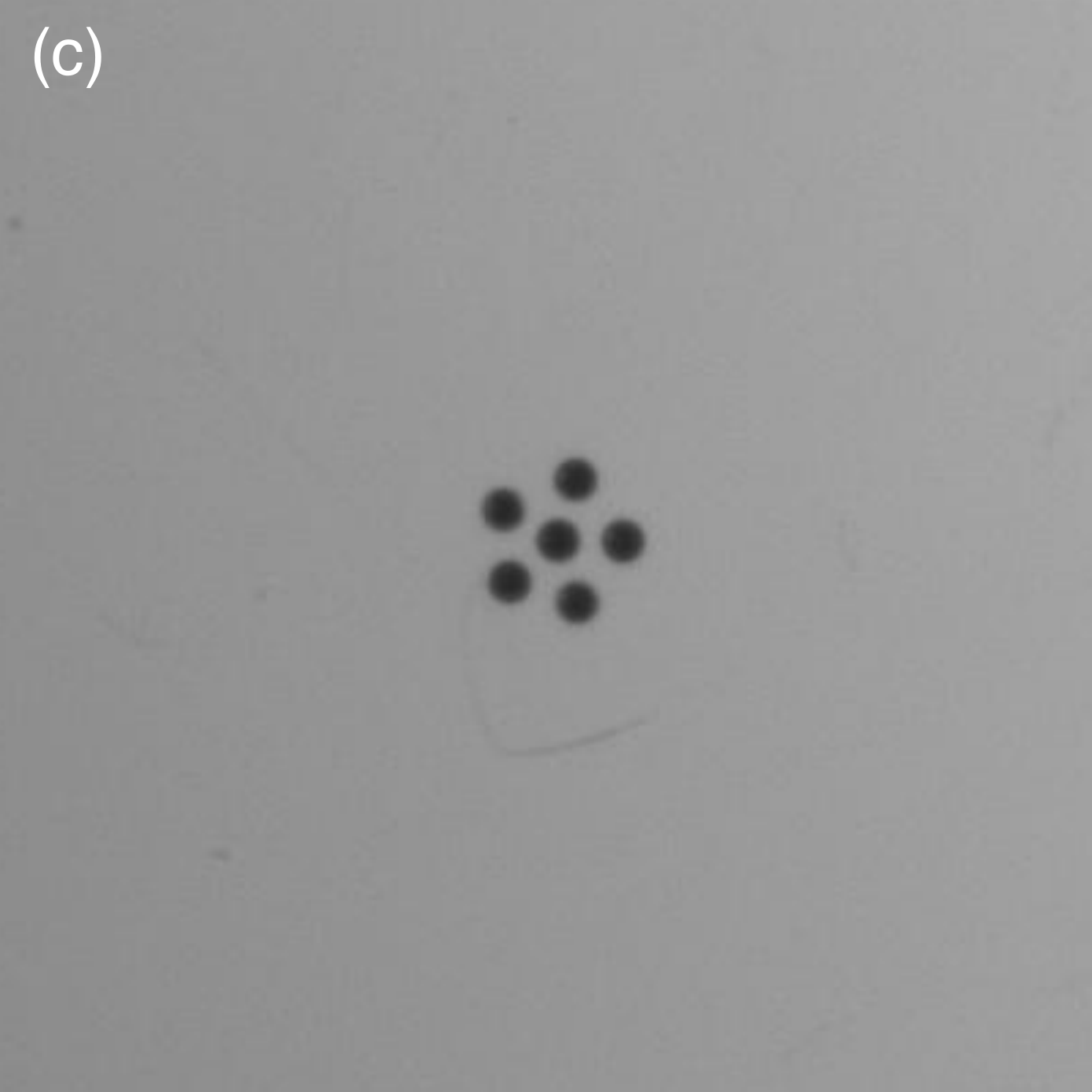} 
\caption{Illustrations of the structures adopted by 500$\mu$m beads once dropped above a water-air interface and submitted to a vertical magnetic field of 27G. One observes the triangular symmetry of the structure.}
\end{figure}

In the \href{}{video}, different experiments are shown. The first set corresponds to beads (D=500$\mu$m) only submitted to a vertical magnetic field. Magnetic repulsion against capillary attraction is illustrated for a field amplitude of 27G. Fig.1 presents snapshots from the \href{}{video}. Fig.1(a),(b) and (c) are respectively for 4, 5 and 6 beads on the surface. One observes a structure with a triangular symmetry. The reason for this symmetry is that the magnetic repulsion and capillary attraction are both central forces.\\

\begin{figure}[!]
\centering
\includegraphics[scale=0.35]{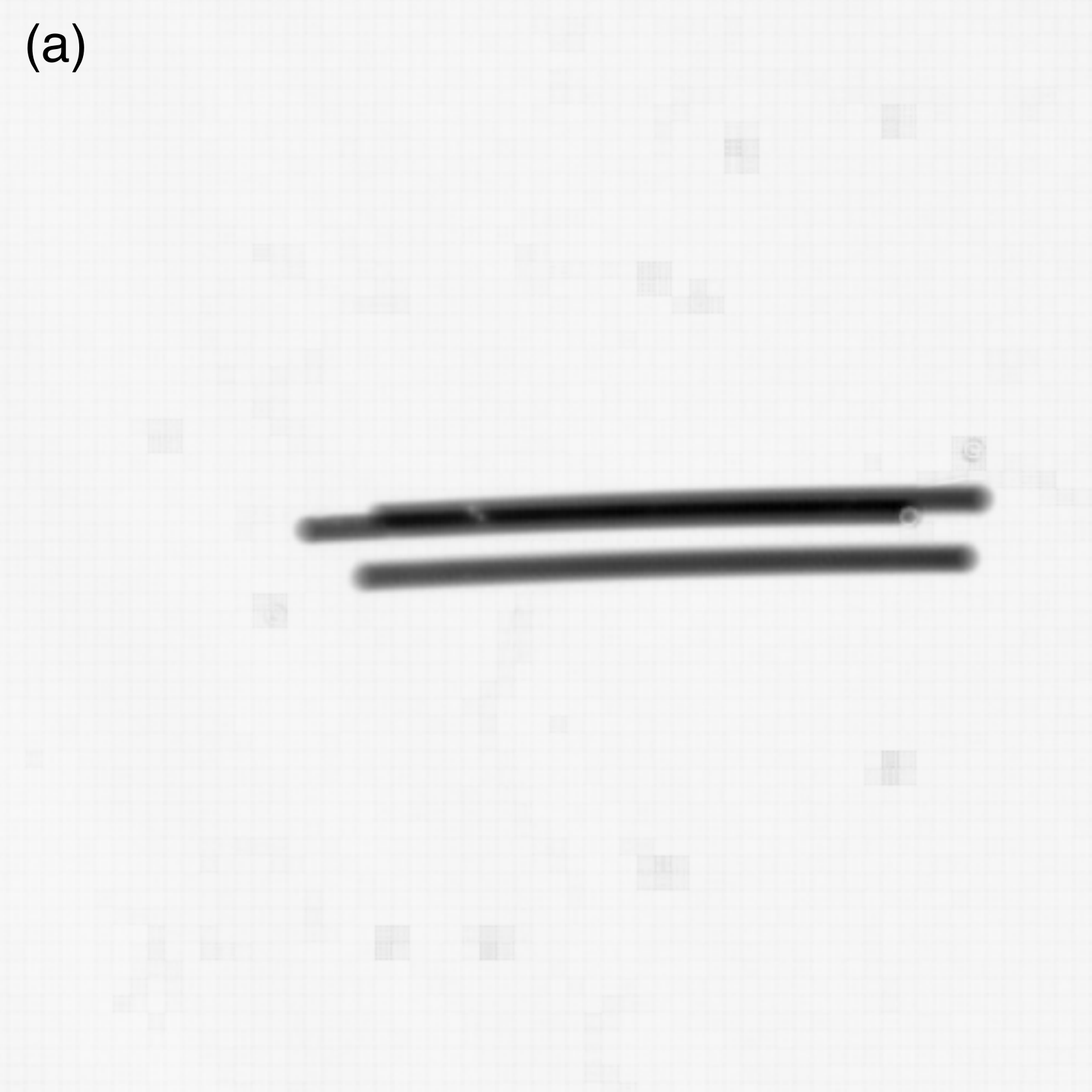} 
\includegraphics[scale=0.35]{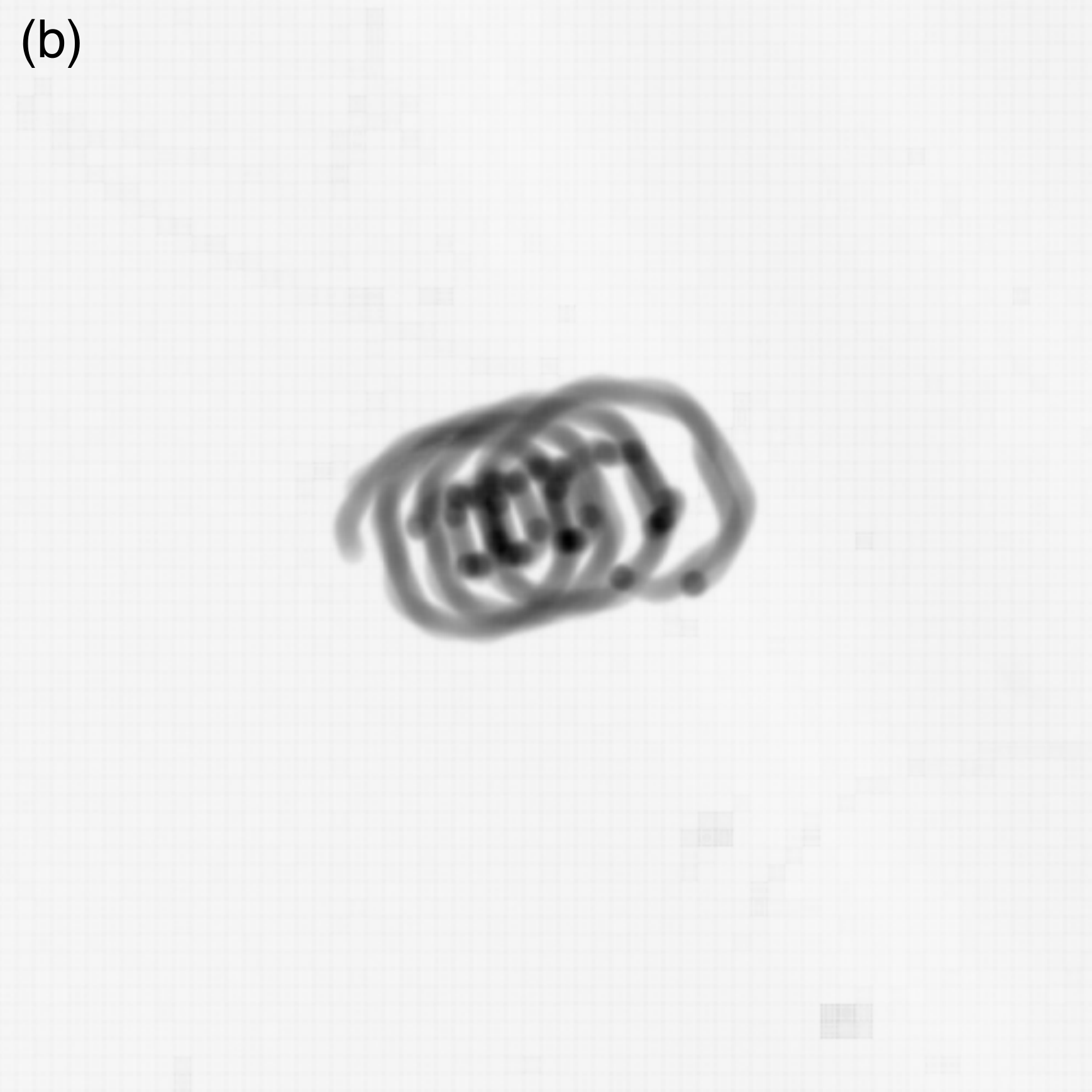} 
\includegraphics[scale=0.35]{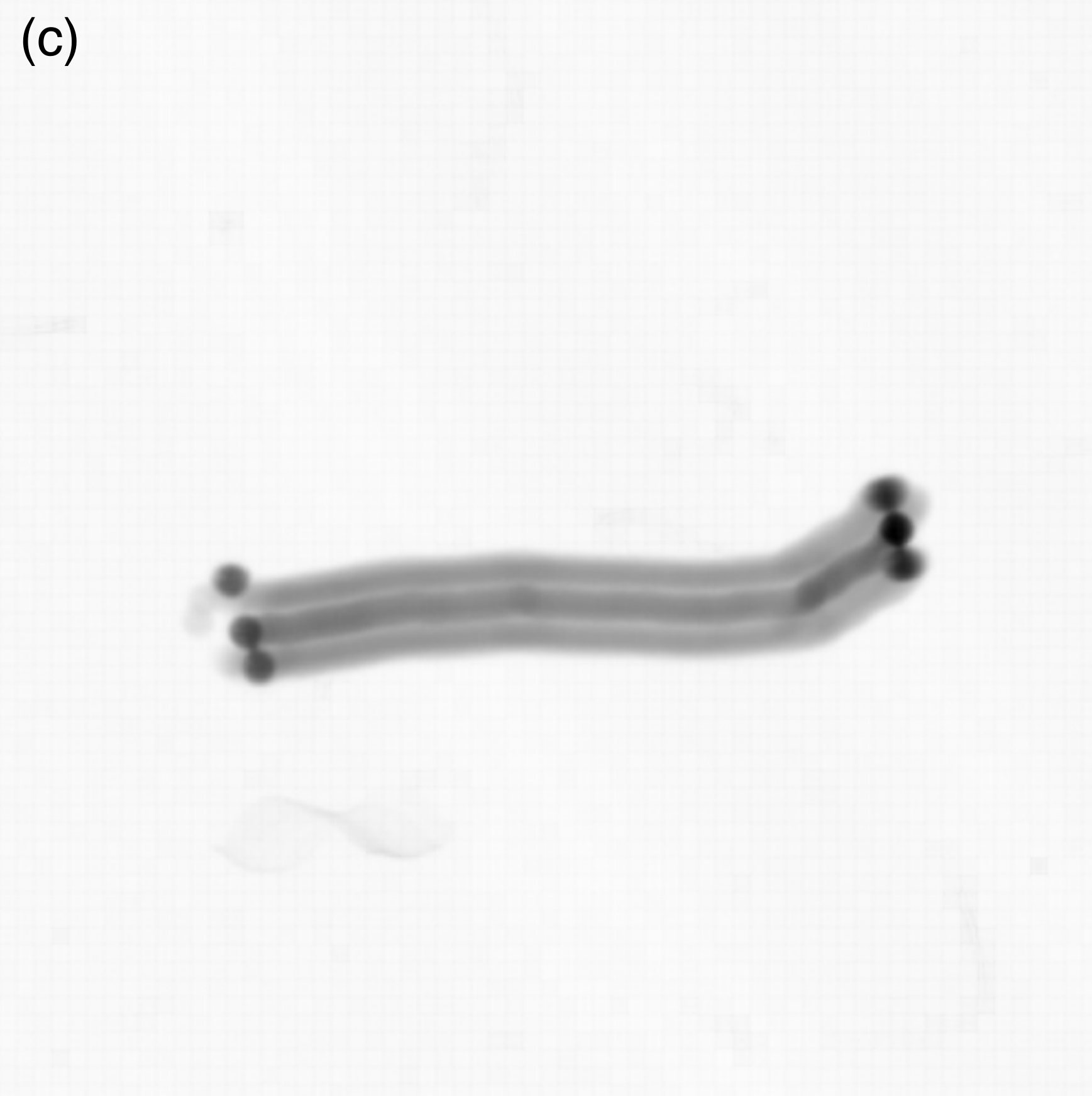} 
\caption{Dynamics of a triplet of 397$\mu$m beads submitted to a vertical constant magnetic field and to an oscillating horizontal magnetic field. Figure (a) corresponds to the \emph{magnetocapillary swimmer}, translating itself along a straight line at the average velocity of 0.2 mm/s. Figure (b) corresponds to the \emph{magnetocapillary rotor}, obtained by changing the initial conditions of the \emph{magnetocapillary swimmer} dynamics. Finally, figure (c) is the \emph{magnetocapillary butterfly},. In this case, the structure collapses because of the strong oscillating field. Nevertheless, one observes that the structure swims.}
\end{figure}

The second set of experiments considers an additional oscillating magnetic field for 397$\mu$m beads. Three dynamics are observed and Fig.2 gives some illustrations. The first dynamics corresponds to the \emph{Magnetocapillary swimmer}, the illustration is shown by Fig.2(a). The vertical field has a intensity of 26G and the horizontal field oscillates at the frequency of 3Hz and has an amplitude of 38G. One observes that the structure deforms and moves along a straight line. It moves over a distance of one centimeter in almost 50 seconds leading to an average velocity of 0.2 mm/s. This dynamics is in total agreement which Purcell's \emph{Scallop theorem} \cite{Purcell}. Indeed, the structure deforms in a non-reciprocal way. This non-reciprocal deformation explains the motion. 

The second dynamics is the \emph{Magnetocapillary rotor} whose behavior is illustrated on Fig.2(b). The parameters in this case are exactly the same as the ones used for the \emph{Magnetocapillary swimmer} but the initial conditions are different. As one observes, the system shifts into another stable orbit where the beads turn around the geometrical center of the structure.  

The final experiment is the \emph{magnetocapillary butterfly}. Fig2(c) gives the corresponding illustration. In this case, the oscillating field is strong enough (Peak-to-peak intensity 32G) to overcome the effect of the vertical field (intensity 26G) leading to a collapse of the structure. Moreover, the high frequency (7.5Hz) make the triplet oscillate quickly resulting in a swimming motion different from the case of the \emph{magnetocapillary swimmer}. By switching off the field and then switching it back on, this swimmer can explore others directions.\\

As a conclusion, this experiment leads to a large collection of surprising behaviors. Triplets of beads may either swim or rotate because of magnetic fields and non-reciprocal deformations.


\begin{thebibliography}{5}
\bibitem{Whitesides} G. M. Whitesides and B. Grzybowski, ``Self-Assembly at All Scales",Soft Matter \textbf{102}, 2418 (2002) 
\bibitem{Vella} D. Vella and L. Mahadevan,``The Cheerios Effect",  Am. J. Phys.  \textbf{73}, 817 (2005) 
\bibitem{Vandewalle} N. Vandewalle, L. Clermont, D. Terwagne, S. Dorbolo, E. Mersch and G. Lumay, ``Symmetry breaking in a few-body system with magnetocapillary interactions", Phys. Rev. E \textbf{85}, 041402 (2012) 
\bibitem{Lumay} G. Lumay, N. Obara, F. Weyer and N. Vandewalle,``Self-assembled magnetocapillary swimmers",  Soft Matter \textbf{9}, 2420 (2013) 
\bibitem{Purcell} E. M. Purcell,``Life at low Reynolds number",  Am. J. Phys. \textbf{45}, 3 (1977) 
\end{thebibliography}
\end{document}